\documentclass[manuscript]{emulateapj}
\usepackage{apjfonts}
\usepackage{epsf}
\usepackage{color}

\newcommand{\HI}{\ion{H}{1}}
\newcommand{\Halpha}{H$\alpha$}
\newcommand{\Ha}{H$\alpha$}
\newcommand{\etal}{{et al.}}
\newcommand{\kms}{km s$^{-1}$}

\font\mitt=cmmi9 scaled\magstep 1
\def\varv{\hbox{\mitt v}}

\shorttitle{Testing MOND in Dwarf and LSB Galaxies}

\shortauthors{Swaters, Sanders, \& McGaugh}

\begin{document}

\title{Testing Modified Newtonian Dynamics with Rotation Curves
of Dwarf and Low Surface Brightness Galaxies}

\author{R. A. Swaters\altaffilmark{1,2}}
\affil{Department of Astronomy, University of
  Maryland, College Park, MD 20742-2421}
\altaffiltext{1}{Department of Physics and Astronomy, Johns Hopkins University, 3400 N. Charles Str., Baltimore, MD 21218}
\altaffiltext{2}{Space Telescope Science Institute, 3700 San Martin Dr., Baltimore, MD 21218}
\email{rob@swaters.net}

\author{R. H. Sanders}
\affil{Kapteyn Institute, P.O. Box 800, 9700 AV Groningen,
  the Netherlands}

\author{S. S.  McGaugh} \affil{Department of Astronomy, University of
  Maryland, College Park, MD 20742-2421}

\begin{abstract}

\noindent Dwarf and low surface brightness galaxies are ideal objects
to test modified Newtonian dynamics (MOND), because in most of these
galaxies the accelerations fall below the threshold below where MOND
supposedly applies. We have selected from the literature a sample of
27 dwarf and low surface brightness galaxies. MOND is successful in
explaining the general shape of the observed rotation curves for
roughly three quarters of the galaxies in the sample presented
here. However, for the remaining quarter, MOND does not adequately
explain the observed rotation curves.  Considering the uncertainties
in distances and inclinations for the galaxies in our sample, a small
fraction of poor MOND predictions is expected and is not necessarily a
problem for MOND.  We have also made fits taking the MOND acceleration
constant, $a_0$, as a free parameter in order to identify any
systematic trends.  We find that there appears to be a correlation
between central surface brightness and the best-fit value of $a_0$, in
the sense that lower surface brightness galaxies tend to have lower
$a_0$. However, this correlation depends strongly on a small number of
galaxies whose rotation curves might be uncertain due to either bars
or warps. Without these galaxies, there is less evidence of a trend,
but the average value we find for $a_0\approx 0.7 \times 10^{-8}$ cm
s$^{-2}$ is somewhat lower than derived from previous studies. Such
lower fitted values of $a_0$ could occur if external gravitational
fields are important.

\end{abstract}


\section{Introduction}

\noindent Modified Newtonian dynamics (MOND) was proposed by Milgrom
(1983a,b) as an alternative to dark matter.  MOND posits that the
effective gravitational force deviates from the Newtonian force: at
accelerations below a critical value, $a_0$, the gravitational force
is proportional to the square root of the Newtonian force; at higher
accelerations, the force is Newtonian.  In principle, MOND predicts
the shape and amplitude of the observed rotation curves from the
observed mass distribution, i.e., gas and stars, with only the
mass-to-light ratio (M/L) of the stellar disk as adjustable parameter.

The rotation curves of a large number of spiral galaxies have been
considered in the context of MOND. For example, Begeman et
al.\ (1991), Sanders (1996), Sanders \& Verheijen (1998), and Sanders
\& Noordermeer (2007) demonstrated that for the spiral galaxies the
observed rotation curves were, in most cases, predicted in detail,
using the MOND prescription, from the observed light and gas
distributions.  Moreover, the M/Ls derived from these fits are
generally astrophysically plausible and consistent with stellar
population synthesis models.  In a handful of these galaxies the
rotation curves predicted by MOND are noticeably different from the
observed rotation curves. These were not considered falsifications of
MOND because for most of these galaxies the authors identified an
obvious problem with the observed rotation curve (e.g., uncertain
inclination or distance) or with the use of the rotation curve as a
tracer of the gravitational force (e.g., disturbed velocity fields).

Although MOND appears to explain the magnitude of the discrepancy in
spiral galaxies, it has has long been known that the theory predicts
more matter than is observed in clusters of galaxies (e.g., Sanders
1999).  In this sense the famous "bullet cluster", which has been
presented as direct emprical evidence for the existence of dark matter
(e.g., Clowe et al.\ 2006), does not pose a new problem for MOND.
Additional undetected matter in some dissipationless form does seem to
be required in clusters in the context of MOND.  In any case,
additional required matter is not formally a falsification of MOND and
in no sense detracts from the success of this algorithm on the scale
of galaxies.

\begin{deluxetable*}{lcrrrrrrrr}[th]
\tabletypesize{\scriptsize}
\tablecaption{Galaxy properties\label{tabsample}}
\tablewidth{0pt}
\tablehead{
\colhead{Name} & \colhead{Source} & \colhead{$\mathrm{D}_a$} &
\colhead{$\mu_0^R$} & \colhead{$h$} & \colhead{$M_R$} &
\colhead{$M_\mathrm{HI}$} & \colhead{$i$} & \colhead{$v_\mathrm{rot}$} &
\colhead{$a_\mathrm{lmp}/a_0$} \\
\colhead{} & \colhead{} & \colhead{Mpc} & \colhead{mag$ \prime\prime^{-2}$} &
\colhead{kpc} & \colhead{mag} & \colhead{$10^8 M_\odot$} & 
\colhead{$^\circ$} & \colhead{km s$^{-1}$} & \colhead{} \\
\colhead{(1)} & \colhead{(2)} & \colhead{(3)} & \colhead{(4)} & \colhead{(5)} &
\colhead{(6)} & \colhead{(7)} & \colhead{(8)} & \colhead{(9)} & \colhead{(10)} \\
}
\startdata
UGC 731 & ~SMBB ~& 8.0 &23.0 &1.65 &-16.6 &7.37 &57 &  74 & 0.25  \\
UGC 3371 & ~SVBA ~&12.8 &23.3 &3.09 &-17.7 &12.2 &49 &  86 & 0.23  \\
UGC 4173 & ~dBB ~&18.0 &24.3 &4.77 &-18.0 &24.3 &40 &  57 & 0.08  \\
UGC 4325 & ~SMBB ~&10.8 &21.6 &1.74 &-18.2 &8.61 &41 &  93 & 0.44  \\
UGC 4499 & ~SMBB ~&13.9 &21.5 &1.59 &-17.9 &13.6 &50 &  74 & 0.20  \\
UGC 5005 & ~dBB ~&  56 &22.9 &4.71 &-18.8 &33.2 &41 &  99 & 0.11  \\
UGC 5414 & ~SSAH ~&10.7 &21.8 &1.59 &-17.7 &7.42 &55 &  61 & 0.26  \\
UGC 5721 & ~SMBB ~& 6.6 &20.2 &0.45 &-16.6 &6.63 &61 &  79 & 0.28  \\
UGC 5750 & ~dBMR ~&  60 &22.6 &5.99 &-19.6 &11.4 &64 &  79 & 0.09  \\
UGC 6446 & ~SSAH ~&12.8 &21.4 &2.00 &-18.5 &15.4 &52 &  80 & 0.20  \\
UGC 7232 & ~SSAH ~& 3.5 &20.2 &0.33 &-15.3 &0.71 &59 &  44 & 0.62  \\
UGC 7323 & ~SSAH ~& 8.7 &21.2 &2.35 &-19.0 &8.48 &47 &  86 & 0.38  \\
UGC 7399 & ~SSAH ~& 8.4 &20.7 &0.79 &-17.1 &7.40 &55 & 109 & 0.35  \\
UGC 7524 & ~SSAH ~& 4.6\rlap{$^\ast$} &22.2 &3.37 &-18.7 &16.7 &46 &  84 & 0.22  \\
UGC 7559 & ~SSAH ~& 4.9\rlap{$^\ast$} &23.8 &1.02 &-14.6 &1.71 &61 &  33 & 0.11  \\
UGC 7577 & ~SSAH ~& 2.5\rlap{$^{\ast\ast}$} &22.5 &0.60 &-14.9 &0.42 &63 &  18 & 0.06  \\
UGC 7603 & ~dBB ~& 7.3 &20.8 &0.96 &-17.0 &6.13 &78 &  64 & 0.21  \\
UGC 8490 & ~SMBB ~& 4.6\rlap{$^\ast$} &20.5 &0.62 &-17.1 &7.01 &50 &  80 & 0.21  \\
UGC 9211 & ~dBB ~&12.6 &22.6 &1.32 &-16.2 &10.5 &44 &  66 & 0.17  \\
UGC 11707 & ~SMBB ~&17.0 &23.1 &4.61 &-18.8 &42.7 &68 & 100 & 0.20  \\
UGC 11861 & ~SMBB ~&  27 &21.4 &6.48 &-20.9 &81.7 &50 & 153 & 0.43  \\
UGC 12060 & ~SSAH ~&16.8 &21.6 &1.89 &-18.1 &20.7 &40 &  75 & 0.16  \\
UGC 12632 & ~SSAH ~& 6.9 &23.5 &2.57 &-17.1 &8.66 &46 &  76 & 0.22  \\
F568-V1 & ~SMBB ~&  86 &22.8 &3.42 &-18.8 &28.1 &40 & 118 & 0.24  \\
F574-1 & ~SMBB ~& 103 &22.6 &4.60 &-19.3 &39.7 &65 & 100 & 0.22  \\
F583-1 & ~dBMR ~&  34 &22.0 &1.71 &-17.5 &19.9 &63 &  84 & 0.15  \\
F583-4 & ~dBMR ~&  52 &22.4 &2.89 &-17.8 &6.30 &55 &  70 & 0.22  \\
\enddata
\tablecomments{(1) the name of the galaxy, (2) source of the rotation
  curve: de Blok \etal\ (2001, dBMR), de Blok \& Bosma (2002; dBB),
  Swaters \etal\ (2003a; SMBB), Swaters \etal\ (2003b; SVBA), Swaters
  \etal\ (2009; SSAH), (3) adopted distance, taken from Swaters \&
  Balcells (2002), except that distances for galaxies marked with
  $^\ast$ were taken from Karachentev et al.\ (2003), those marked
  with $^{\ast\ast}$ were taken from M\'endez et al. (2002) (4--6)
  surface brightness, scale length, and absolute magnitude, from
  Swaters \& Balcells (2002), de Blok et al.\ (1995), and McGaugh \&
  Bothun (1994), (7) HI mass, from Swaters et al.\ (2003), de Blok et
  al.\ (1996), and van der Hulst et al.\ (1993) (8) inclinations, from
  Swaters et al.\ (2004) and de Blok et al.\ (1996), (9) maximum
  rotation velocity, (10) acceleration at the last measured point of
  the rotation curve.}
\end{deluxetable*}

MOND is more strongly tested by spiral galaxies with well-determined
distances because the $a_0$ depends inversely upon distance.  Bottema
et al.\ (2002) considered a sample of 4 spiral galaxies with
well-measured rotation curves and Cepheid-based distances determined
as part of the HST program on the extragalactic distance scale (Sakai
et al.\ 2000). Two galaxies are in good agreement with MOND
predictions, but for NGC~2841 and NGC~3198 the rotation curves are not
in agreement with MOND, unless the distance to NGC~2841 is at least
20\% larger, and the distance to NGC~3198 is at least 10\% smaller
than their Cepheid distances indicate. Bottema et al.\ (2002) argued
that NGC~3198 is probably consistent with MOND within the
uncertainties of the Cepheid distance. With respect to NGC 2841,
Milgrom and Sanders (2008) have demonstrated that a reasonable match
to the observed rotation curve can be achieved with alternative forms
of the function that described the transition from the Newtonian to
the MOND regime.

Dwarf and low surface brightness (LSB) galaxies provide an especially
good test for MOND, because in most of these galaxies the acceleration
is below the MOND threshold of $a_0 \approx 10^{-8}$~cm~s$^{-2}$
(determined empirically from rotation curve fits, see e.g., Sanders \&
McGaugh 2002) at all radii, whereas in high surface brightness spiral
galaxies the acceleration is usually above the threshold in the inner
parts.  Milgrom (1983b) predicted that LSB galaxies should exhibit a
large discrepancy between the detectable and Newtonian dynamical mass
within the optical disk, and this prediction has since been confirmed
(e.g., de Blok \& McGaugh 1997; McGaugh \& de Blok 1998; Swaters
\etal\ 2000).

Apart from this general prediction, the MOND prescription is successful
in predicting the detailed rotation curves of LSB galaxies as well. De
Blok \& McGaugh (1998) found that for their sample of 15 LSB galaxies
all rotation curves are well fitted with MOND, although for 6 galaxies
modest adjustments of the inclination were necessary in order to get
an acceptable match to the rotation curve.  The sample presented in
Sanders \& Verheijen (1998) contains 12 LSB galaxies, and most of
these the rotation curves can be well explained by MOND as well.

\begin{figure*}[p]
\resizebox{0.9\hsize}{!}{\includegraphics{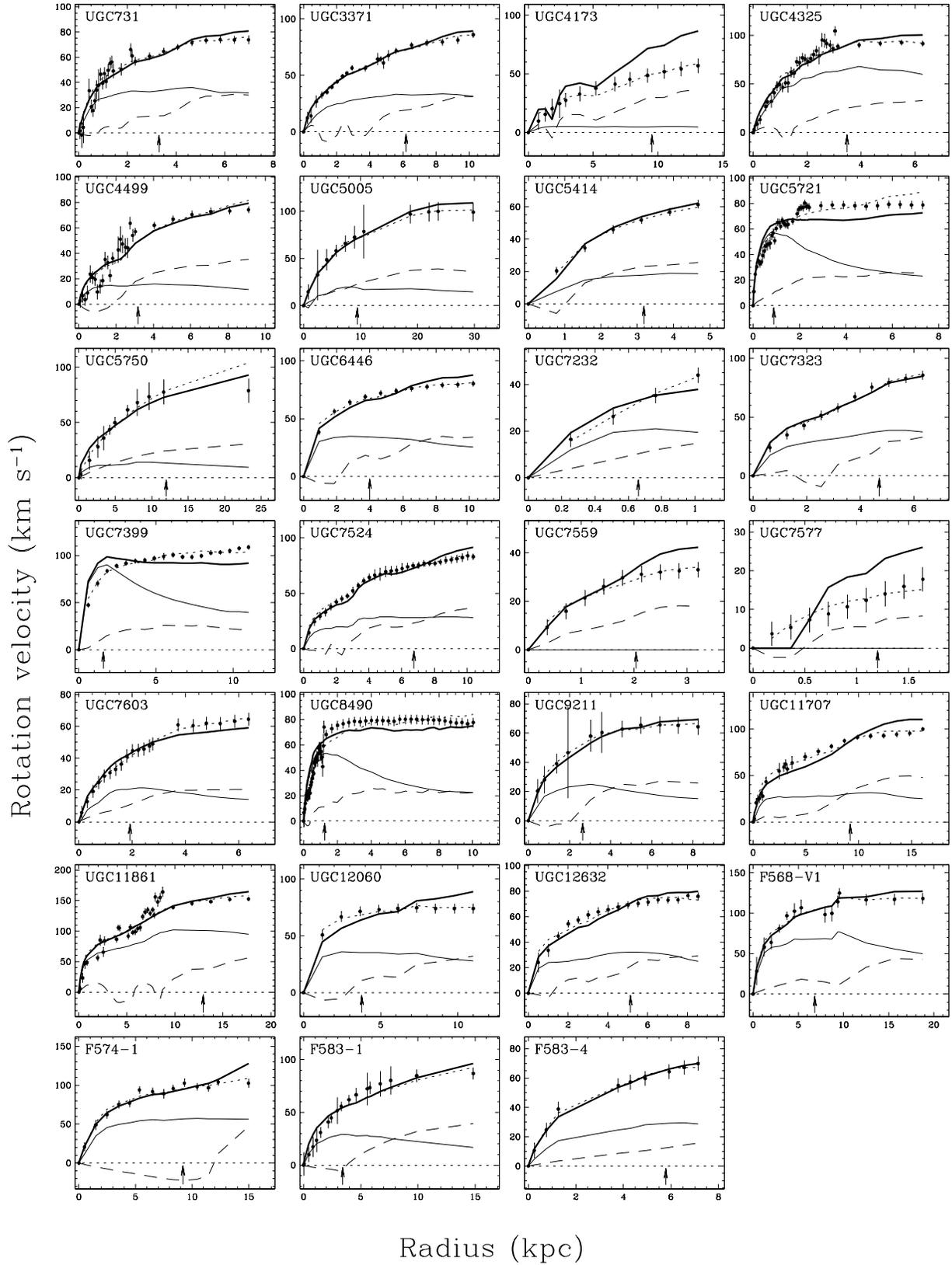}}
\caption{MOND predictions. The dots represent the measured rotation
  velocities and their errors. The dashed line represents the
  contribution of the HI to the rotation curve. In cases where there
  is little or no HI at small radii in the HI disks, the HI at larger
  radii produces a net outward gravitational force, which is
  represented here by the negative velocities. The thin solid line
  represents the contribution of the stars. The thick solid line is
  the best MOND fit with only the stellar mass-to-light ratio as a
  free parameter. The dotted line is the best MOND fit with both the
  mass-to-light ratio and the distance as free parameter.
\label{figfits}}
\end{figure*}

The results for dwarf galaxies show a fairly large spread.  Milgrom \&
Braun (1988) concluded that the observed rotation curve of DDO~154 is
in very good agreement with the rotation curve predicted by MOND.
Lake (1989) considered a sample of 6 dwarf galaxies in the context of
MOND and concluded that acceptable fits require systematically lower
values for $a_0$ than were found for spiral galaxies.  In
addition, Lake (1989) found an apparent correlation between the
asymptotic rotation velocity and the fitted value of $a_0$.  Milgrom
(1991), however, pointed out that uncertainties in inclinations and
distances can well explain the fact that MOND appears to fail for
these galaxies. The rotation curves of the dwarf galaxies included in
the samples of Begeman et al.\ (1991) and Sanders (1996) also agree
well with those calculated with MOND. On the other hand,
Blais-Ouellette et al.  (2001) reported that MOND provides poor
predictions of the rotation curves of the three dwarf galaxies in
their sample, and reported that a higher value of $a_0$ is needed to
make their rotation curves derived from Fabry-Perot observations
consistent with MOND. More recently, Milgrom \& Sanders (2007)
concluded that the rotation curves of the four low-mass galaxies in
their sample are explain well in the context of MOND.

Given this apparent spread in conclusions on whether MOND correctly
predicts the rotation curves of dwarf galaxies, we here test MOND
against a sample of rotation curves of dwarf and LSB galaxies that
have recently become available in the literature. We included
both dwarf and LSB galaxies because the physical properties of these
galaxies are similar, although some dwarf galaxies have high surface
brightness, and some LSB galaxies are large in size.

There are several difficulties inherent to such a sample. For example,
because most dwarf galaxies are nearby, different methods to estimate
the distance have been used, and this may introduce a dispersion in
the derived parameters.  Inclination is another uncertainty for a
class of objects which are known to be irregular with, in some cases,
large scale asymmetries.  Moreover, the assumption that the rotation
curve is a tracer of the radial force distribution may, itself, be
questionable in cases where asymmetries are present.  Bearing these
considerations in mind, one might not expect, a priori, the agreement
between the predicted and observed rotation curves to be as good as
noted for earlier samples. A systematic failure, however, would be
problematic for MOND.

The outline of this paper is as follows.  In Section~\ref{secsample}
we describe the sample used in this paper. In Section~\ref{secfitproc}
we describe the fitting procedure, and in Section~\ref{secfitres} the
fitting results. Next, we comment on individual galaxies in
Section~\ref{secindiv}.  We describe the uncertainties in
Section~\ref{secuncert}, and we discuss the results in
Section~\ref{secdisc}. Finally, we present our conclusions in
Section~\ref{secconc}.

\section{The sample}
\label{secsample}

\noindent The sample we present here is compiled from data from five
studies. A large fraction comes from the \HI\ rotation curves of a
sample of 62 dwarf galaxies (Swaters 1999; Swaters \etal\ 2009,
hereafter SSAH). The resolution of the HI data used to derive the
rotation curves is $30''$. Because of this relatively low resolution,
the data have been corrected for beam smearing as described in Swaters
(1999) and SSAH.  From this original sample of 62 dwarf galaxies, we
selected galaxies with inclination in the range $40^\circ\le i\le
80^\circ$, and we only included rotation curves that were classified
by SSAH as high quality.  Galaxies classified by SSAH as having lower
quality either have a signal-to-noise ratio that is too low, or are
too asymmetric to derive a reliable rotation curve. To avoid the
uncertainties associated with these low quality rotation curves, we
have excluded these from our sample.  The resulting sample contains 19
rotation curves.

The second source is the sample of \Ha\ rotation curves presented in
Swaters \etal\ (2003a). They obtained high resolution rotation curves
from \Ha\ long-slit spectroscopy for a sample of 10 dwarf that are
part of the SSAH sample described above, and 5 LSB galaxies from the
sample presented in de Blok \etal\ (1996). As
described in Swaters et al.\ (2003a), these \Halpha\ rotation curves
have been combined with the \HI\ data presented SSAH and de Blok et
al.\ (1996). Galaxies with $i<40^\circ$ or $i>80^\circ$ were not
included in the sample here. F568-3 was also excluded because of its
strong bar.

The third source is the sample of LSB galaxies presented in de Blok,
\etal\ (2001). They present long-slit \Ha\ rotation curves for a
sample of 30 LSB galaxies. From their sample we have selected those
galaxies for which optical photometry and \HI\ imaging are available
and with inclinations in the range $40^\circ\le i\le 80^\circ$.

The fourth source is the sample presented in de Blok \& Bosma (2002),
consisting of \HI-\Halpha\ hybrid rotation curves. From this sample we
have selected the galaxies that overlap with the SSAH sample. We
excluded UGC~3851 because its rotation curve depends heavily on the
uncertain \HI\ rotation curve as derived by SSAH and which they deemed
to be of low quality.  UGC~7524 was also excluded because this galaxy
is well resolved by the \HI\ observations and the rotation curve
derived from the two-dimensional data probably gives a better
representation of the gravitational potential than the one derived
from long-slit data. The rotation curves presented in de Blok \& Bosma
(2002) have been combined with the \HI\ rotation curves presented in
SSAH as described in de Blok \& Bosma (2002).

The fifth and final source is the rotation curve of UGC~3371 presented
in Swaters et al.\ (2003b). They obtained high-resolution,
two-dimensional \Halpha\ data for this galaxy.

Where the samples are overlapping, we have used them in this order of
decreasing priority: Swaters et al. (2003b), Swaters et al. (2003a),
de Blok \& Bosma (2002), de Blok et al. (2001), SSAH.  Most rotation
curves in our sample have been derived from \Halpha\ data for the
central parts and \HI\ data for the outer parts. However, for the 10
galaxies that came from SSAH (see column~2 in Table~\ref{tabsample})
the rotation curves are entirely based on \HI\ data.  We have not
corrected the rotation curves in these samples for pressure support,
because these corrections tend to be small in comparison to the
rotation velocities, and probably uncertain (see also SSAH).

The combined sample does not constitute a complete sample of dwarf and
LSB galaxies. Nonetheless, this sample contains galaxies with a wide
range in luminosity and surface brightness for which high-quality
rotation curve are available. This sample is therefore well suited to
test MOND in late-type dwarf and LSB galaxies.

Because many of the galaxies in our sample are at small distances,
their systemic velocities are a poor indicator of their distance.
Where possible, we use other distance indicators, based on the
compilation of distances found in the literature presented in Swaters
\& Balcells (2002), supplemented with other recent distance estimates
based on the tip of the red giant branch (M\'endez et al.\ 2002;
Karachentsev et al.\ 2003). In order of decreasing priority, the
distances presented in Table~\ref{tabsample} are based on Cepheids,
the tip of the red giant branch, brightest stars, group membership,
and the recession velocities corrected for Virgocentric flow as
described in Swaters \& Balcells (2002). For a more complete
discussion on the distances to these dwarf galaxies, see Swaters \&
Balcells (2002). The galaxies in our sample that are not in the SSAH
sample are all at large distances, and for these the distances have
been calculated based on a Hubble constant of 70 \kms\ Mpc$^{-1}$,
following Sanders \& McGaugh (2002).

\section{Fitting procedure}
\label{secfitproc}

\noindent To calculate the MOND rotation curves, we have used the same
method as used in Sanders (1996) and Sanders \& Verheijen (1998). To
calculate the contribution of the stellar component to the disk, we
have assumed that the stellar mass resides in an infinitely thin disk,
that the $R$-band light accurately traces the mass distribution of the
stellar component, and that the stellar mass-to-light ratio is
constant with radius.  The gas mass distribution is also assumed to be
in a thin disk. We assumed that the \HI\ is optically thin, and we
corrected for the mass fraction of helium by scaling the \HI\ mass by
a factor of 1.32. The optical radial profiles were taken from Swaters
\& Balcells (2002), de Blok et al.\ (1995) and McGaugh \& Bothun
(1994). The \HI\ radial profiles were taken from Swaters et
al. (2002), de Blok et al.\ (1996), and van der Hulst et
al.\ (1993). For the galaxies in which the \HI\ distribution is poorly
resolved, we calculated the radial profiles following the algorithm
described by Warmels (1988) that uses a iterative deconvolution scheme
(Lucy 1974) to correct for the effects of beam smearing (see Swaters
et al.\ 2002 for more details).  For both the stellar and the gaseous
disk, we assumed that the mass distribution is well represented by the
radial average.

To calculate the MOND rotation curves of the visible components,
ideally the MOND field equation of Bekenstein \& Milgrom (1984) should
be used. However, this is computationally expensive, and Milgrom
(1986) has shown that the differences between the results derived for
the field equation and those from the original MOND prescription are
usually much smaller than, and practically never larger than
5\%. Therefore, we apply the usual MOND formula:
\begin{equation}
\mu(g_\mathrm{M}/a_0){\bf g}_\mathrm{M} = {\bf g}_\mathrm{N},
\label{eqmond}
\end{equation}
where $a_0$ is the MOND acceleration parameter, ${\bf g}_M$ is the
MOND acceleration, ${\bf g}_N$ is the Newtonian acceleration, and
\begin{equation}
\mu(x) = x(1+x^2)^{-1/2},
\label{eqmu}
\end{equation}
which is the commonly assumed form having the required asymptotic
behaviour. For $g\gg a_0$ gravity is Newtonian, and for $g\ll a_0$
gravity is of the MOND form with $g=(g_Na_0)^{1/2}$.  A little algebra
shows that the rotation curve for MOND can be expressed as:
\begin{equation}
\varv_{rot}^2 = {\varv_{sum}^2\over{\sqrt{2}}}\sqrt{1+\sqrt{1+(2ra_0/{\varv_{sum}^2)^2}}}.
\label{eqfit}
\end{equation}
Here, $r$ is the radius and
\begin{equation}
\varv_{sum}^2 = \Upsilon_{\ast,d} \varv_d^2+\Upsilon_{\ast,b} \varv_b^2+\varv_g^2,
\label{eqsum}
\end{equation}
with $\varv_d$, $\varv_b$ and $\varv_g$ the contribution of the stellar disk,
bulge and gas to the rotation curve, respectively, calculated in the
Newtonian regime as described above, and $\Upsilon_{\ast,d}$ and
$\Upsilon_{\ast,b}$ are the stellar mass-to-light ratios of the disk
and the bulge. Eq.~\ref{eqfit} is fit to the observed rotation curve
with a least-squares algorithm. None of the galaxies in our sample
have significant bulges, which eliminates $\Upsilon_{\ast,b}$ as a
free parameter.

It should be noted that in the context of MOND the internal dynamics
of a system is affected by the external acceleration field. If the
external field becomes comparable to or larger than $a_0$, a galaxy in
this external field will always be in the Newtonian regime, even if
the internal accelerations are low (Milgrom 1983a,b).

\section{Results}
\label{secfitres}

\noindent We made MOND fits by fitting Eq.~\ref{eqfit} to the observed
rotation curves with a least squares algorithm.  Because $a_0$ is a
universal constant in MOND, we first made fits with only
$\Upsilon_{\ast,d}$ as a free parameter.  Following e.g., Sanders \&
McGaugh (2002), the acceleration parameter was fixed at $a_0=1.0\times
10^{-8}$ cm~s$^{-2}$ ($=3080$ km$^2$ s$^{-2}$ kpc$^{-1}$).

In addition to the one parameter fits, we have also made MOND fits
with the distance and inclination as free parameters in order to
investigate the effects of uncertainties in these parameters. In
addition, we made fits with $a_0$ as a free parameter to measure $a_0$
and look for trends.

\bigskip
\subsection{One-parameter MOND fits}

\begin{figure}
\plotone{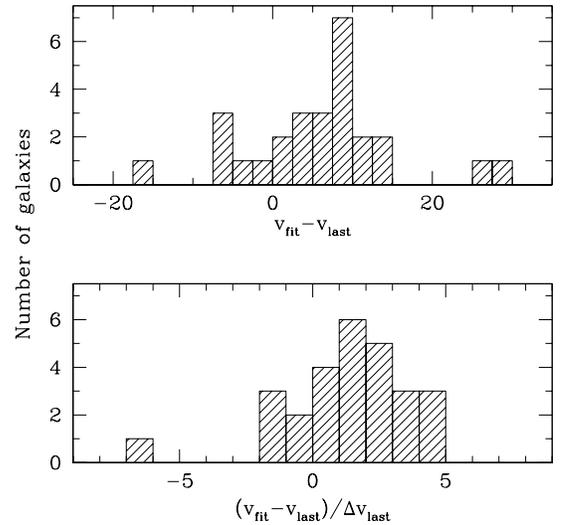}
\caption{Histogram of the difference between the MOND-predicted
  rotation velocity and the observed velocity at the last measured
  point of the rotation curve (top panel). The bottom panel shows the
  same difference, but normalized by the uncertainty in the observed
  rotation velocity at the last measured point.\label{fighistouter}}
\end{figure}

\begin{figure}
\plotone{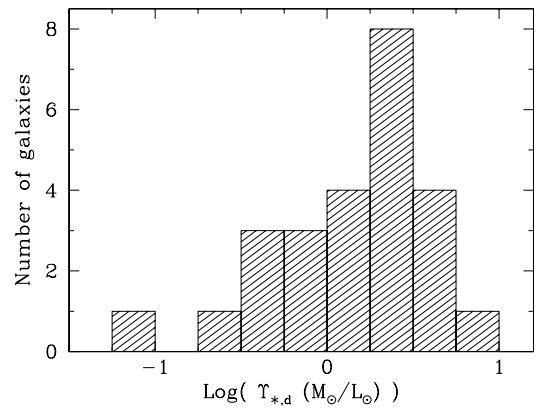}
\caption{Histogram of best fit $\Upsilon_{\ast,d}$ values for models
  with only $\Upsilon_{\ast,d}$ as a free parameter. Two galaxies with
  best-fit $\Upsilon_{\ast,d}=0$ are not shown.\label{fighistmlfix}}
\end{figure}

\noindent The one-parameter MOND fits, with only $\Upsilon_{\ast,d}$
as a free parameter, are shown in Figure~\ref{figfits}. In each panel
of this figure, the dots represent the measured rotation velocities
and their uncertainties, the thin solid line shows the contribution of
the stellar disk, and the dashed line the contribution of the HI. The
best fitting MOND rotation curve is indicated by the thick solid line.

A first inspection of the MOND fits presented in Figure~\ref{figfits}
shows that roughly three quarters of the rotation curves are well fit
by MOND, with only $\Upsilon_{\ast,d}$ as a free parameter. This is
perhaps surprising, given the possible uncertainties due to
inclination, distance, and asymmetries. A closer look shows that more
than half of the galaxies deviate somewhat from the observed rotation
rotation in a systematic way: the MOND curve predicts higher rotation
velocities in in the outer regions, and lower in the central
regions. This is seen clearly in e.g., UGC~6446 and UGC~12060, and it
is also visible, although less pronounced, in e.g., UGC~731 and
UGC~3371.

\begin{deluxetable*}{lrrrrrrrrrr}[th]
\tabletypesize{\scriptsize}
\tablecaption{Best fit parameters\label{tabfits}}
\tablewidth{0pt}
\tablehead{
\colhead{} & \multicolumn{1}{c}{$a_0$ fixed} & \multicolumn{3}{c}{$a_0$ free} & \multicolumn{3}{c}{distance free} & \multicolumn{3}{c}{inclination free} \\
\colhead{} & \multicolumn{1}{c}{\hrulefill} & \multicolumn{3}{c}{\hrulefill} & \multicolumn{3}{c}{\hrulefill} & \multicolumn{3}{c}{\hrulefill} \\
\colhead{UGC} & \colhead{$\Upsilon_{R\mathrm{,x}}$} & \colhead{$a_0$} &
\colhead{$\Delta a_0$} & \colhead{$\Upsilon_{R\mathrm{,f}}$} & \colhead{$d$} &
\colhead{$\Delta d$} & \colhead{$\Upsilon_{R\mathrm{,d}}$} & \colhead{$i$} &
\colhead{$\Delta i$} & \colhead{} \\
\colhead{} & \colhead{(M/L)$_\odot$} &
\multicolumn{2}{c}{km$^2$ s$^{-2}$ kpc$^{-1}$} & \colhead{(M/L)$_\odot$} &
\colhead{} & \colhead{} & \colhead{(M/L)$_\odot$} & \colhead{$^\circ$} &
\colhead{$^\circ$} & \colhead{(M/L)$_\odot$} \\
\colhead{(1)} & \colhead{(2)} & \colhead{(3)} & \colhead{(4)} & \colhead{(5)} &
\colhead{(6)} & \colhead{(7)} & \colhead{(8)} & \colhead{(9)} & \colhead{(10)} &
\colhead{(11)} \\
}
\startdata
UGC 731 &  4.3 &1460 & 445 & 8.5 &0.61 &0.15 & 4.6 &38 &5 &10.2  \\
UGC 3371 &  2.8 &2050 & 590 & 4.2 &0.81 &0.06 & 2.9 &41 &5 & 4.3  \\
UGC 4173 &  0.1 &550 & 350 & 0.6 &0.43 &0.08 & 0.1 &24 &5 & 0.7  \\
UGC 4325 &  3.8 &800 & 500 & 6.4 &0.41 &0.15 & 2.5 &24 &5 &12.9  \\
UGC 4499 &  0.3 &3510 & 480 & 0.2 &1.07 &0.07 & 0.3 &54 &6 & 0.2  \\
UGC 5005 &  0.5 &1890 &1100 & 1.3 &0.78 &0.21 & 0.8 &35 &6 & 1.3  \\
UGC 5414 &  0.5 &1880 &1200 & 1.0 &0.78 &0.20 & 0.7 &43 &20 & 1.1  \\
UGC 5721 &  3.9 &9340 & 900 & 1.5 &1.81 &0.09 & 3.9 &69 &3 & 4.4  \\
UGC 5750 &  0.3 &5350 &1600 & 0.0 &1.32 &0.42 & 0.0 &63 &30 & 0.3  \\
UGC 6446 &  1.1 &1710 & 300 & 2.1 &0.69 &0.09 & 1.3 &39 &3 & 2.5  \\
UGC 7232 &  1.3 &15300 &7000 & 0.0 &2.20 &0.57 & 0.1 &63 &22 & 1.2  \\
UGC 7323 &  0.8 &4270 &1740 & 0.5 &1.16 &0.25 & 0.8 &57 &18 & 0.5  \\
UGC 7399 &  9.6 &14200 &1500 & 1.8 &2.21 &0.10 & 7.2 &72 &3 &11.5  \\
UGC 7524 &  0.9 &1290 & 350 & 2.4 &0.61 &0.11 & 1.2 &33 &3 & 3.0  \\
UGC 7559 &  0.0 &650 & 680 & 3.2 &0.48 &0.27 & 0.8 &32 &9 & 3.4  \\
UGC 7577 &  0.0 &80 & 325 & 0.8 &0.22 &0.21 & 0.0 &20 &10 & 0.8  \\
UGC 7603 &  0.8 &6080 &1100 & 0.2 &1.40 &0.13 & 0.4 &63 &10 & 0.3  \\
UGC 8490 &  3.0 &6690 & 700 & 1.3 &1.51 &0.08 & 2.6 &67 &3 & 2.6  \\
UGC 9211 &  2.4 &2110 &1000 & 4.5 &0.82 &0.26 & 3.2 &38 &8 & 4.7  \\
UGC 11707 &  1.2 &1240 & 190 & 3.8 &0.58 &0.07 & 1.9 &86 &2 & 8.9  \\
UGC 11861 &  2.0 &2310 & 500 & 2.5 &0.70 &0.12 & 1.8 &39 &5 & 3.2  \\
UGC 12060 &  1.8 &690 & 270 & 5.6 &0.32 &0.10 & 1.6 &21 &4 &13.7  \\
UGC 12632 &  3.0 &1190 & 390 & 8.0 &0.54 &0.14 & 3.6 &30 &5 &11.4  \\
F568-V1 &  5.4 &1740 & 700 & 8.3 &0.67 &0.20 & 5.1 &31 &5 &10.7  \\
F574-1 &  3.0 &950 & 460 & 6.5 &0.34 &0.14 & 2.1 &90 &2 &47.5  \\
F583-1 &  2.4 &2590 & 990 & 2.9 &0.91 &0.20 & 2.5 &52 &18 & 2.9  \\
F583-4 &  1.9 &610 &3500 & 6.9 &0.24 &0.68 & 1.6 &24 &11 &17.2  \\
\enddata
\end{deluxetable*}

The degree to which MOND appears to predict higher velocities in the
outer rotation curves is shown in Figure~\ref{fighistouter}, in which
a histogram of the difference between the observed and model
velocities is presented, as well as one in which the differences have
been normalized by the uncertainty in the rotation velocity of the
last measured point. In 21 out of 27 galaxies, MOND predicts higher
than observed rotation velocities, whereas it predicts lower
velocities in only 6 cases. The discrepancy is more than twice
the uncertainty in the last measured point in 11 galaxies.

In Figure~\ref{fighistmlfix} we give the histogram of the best fit
values for $\Upsilon_{\ast,d}$. The mass-to-light ratios span a large
range, from 0 up to 10. The average $R$-band $\Upsilon_{\ast,d}$ as
derived from the best fits its $2$ $M_\odot/L_\odot$.

\subsection{MOND fits with distance free}

\noindent Given that some of the adopted distances may be uncertain,
it is possible that, for some of the galaxies for which MOND does not
correctly predict the rotation curves, we have adopted an incorrect
distance.  We will investigate this by leaving the distance as a free
parameter in the fits.

With distance as a free parameter, not only does the radial scale of
the rotation curve change, but at the same time $\varv_d\propto
1/\sqrt{d}$ and $\varv_g\propto\sqrt{d}$, where
$d=\mathrm{D}_\mathrm{MOND}/\mathrm{D}_a$ is the fractional distance
change. Strictly speaking, these relations are only correct if both
the stellar and \HI\ disks are infinitely thin. For disks with finite
thicknesses, the shape of the rotation curves depends on the assumed
thickness of the disk, and hence the shape of the rotation will change
slightly if a difference distance is assumed.

The fits with distance as a free parameter are shown in
Fig.~\ref{figfits} as the dotted lines. From this figure it is clear
that, with distance as a free parameter, good fits are obtained in
virtually all cases. This demonstrates that a different distance can
improve the fit. The best fit values for $d$, the fractional distance
change, are listed in Table~\ref{tabfits}.  The uncertainties on $d$
have been derived from the 68\% confidence levels. However, because
the errors on the rotation curves are non-Gaussian, because \HI\ the
points are correlated, and because the rotation curves and the errors
on the points may be affected by systematic effects, the confidence
levels and the uncertainties derived from them should be considered
estimates.

\begin{figure}
\plotone{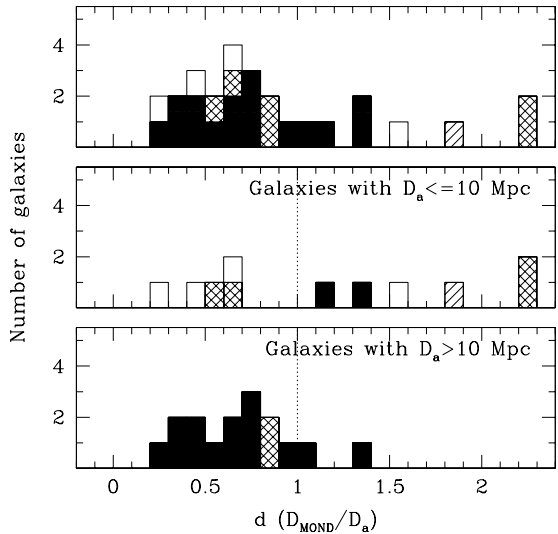}
\caption{Histograms of $d$, the ratio between the best fit distance
  $\mathrm{D}_\mathrm{MOND}$ found from a MOND fit with distance free,
  and $\mathrm{D}_a$, the adopted distance, for all galaxies (top
  panel), galaxies with $D_a<10$ Mpc (middle panel), and for galaxies
  with $D_a>10$ Mpc (bottom panel). The filled areas represent
  galaxies with distances determined from their \HI\ recession
  velocities, the open areas represent galaxies with tip-of-the-RGB
  distances, the hatched areas represent brightest star distances, and
  the cross-hatched areas group membership
  distances.\label{figd0hist}}
\end{figure}

In Fig.~\ref{figd0hist} the distribution of $d$ is shown.  The best
fit fractional distance changes tend to be smaller than 1. As can be
seen in the bottom panel of Fig.~\ref{figd0hist}, this is mostly due
to galaxies at larger distances and for which the distances were
determined from the Hubble flow.  For these galaxies, the average
fractional distance change $\langle d\rangle$ is $0.68$, with a
dispersion of $0.29$. Given that all these galaxies are at relatively
large distances, it is unlikely that the Hubble flow distances are
uncertain by that much, in particular because small $d$ are found even
for galaxies at the largest distances.

For galaxies that are close by ($<10$~Mpc; middle panel of
Fig.~\ref{figd0hist}), the best fit fractional distance changes appear
to scatter around $d=1$ (i.e., no distance change). Such a large
scatter is not unexpected because the galaxies are close-by and
different methods have been employed to determine the distances to
these galaxies.

\subsection{MOND fits with inclination free}
\label{incfits}

\noindent For most galaxies in our sample the inclination cannot be
determined from the kinematics, because the rotation curves for most
of the galaxies in this sample rise slowly without a well-defined flat
part. As a result, for most galaxies in our sample the inclinations
have been estimated from the morphology.  For the galaxies in the SSAH
sample, the inclinations mostly have been determined from the axis
ratios of the outer isophotes (as described by Swaters \& Balcells
2002), except for those galaxies were the velocity field allowed a
reliable estimate (see SSAH for details). For the LSB galaxies, the
inclinations have been taken from de Blok \& McGaugh (1997), who used
the \HI\ maps and optical images to estimate the inclinations.  Given
how these inclinations have been derived, the inclinations of some of
the galaxies in our sample are likely to be uncertain. To investigate
whether some of MOND-predicted rotation curves can be improved by
changing their inclinations, we have made MOND fits with inclination
as a free parameter.

If the inclination is left free in the fits, the amplitude of the
rotation curve changes with $\sin(i_\mathrm{fit})/\sin i$, and at the
same time, the contribution of the \HI\ and the stars to the rotation
curve change with $\cos(i_\mathrm{fit})/\cos i$ (if optically thin).
The latter is only valid for modest changes in the inclination,
especially at high inclinations. When the changes are large, the
rotation curves and density distributions may also change in shape.

In Table~\ref{tabfits} we list the best fit
inclinations. Uncertainties on the inclinations have been estimated
from the 68\% confidence levels in the fit, with the same caveats as
described above for the fits with distance free.  We do not show
the best fits in Fig.~\ref{figfits}, because the fits with
inclinations free are almost indistinguishable from those with
distance as a free parameter, except for UGCs~5721, 7323, 7399, and
8490, for which the fits are nearly indistinguishable from the
original fit with only $\Upsilon_{\ast,d}$ as a free parameter.

\begin{figure}
\plotone{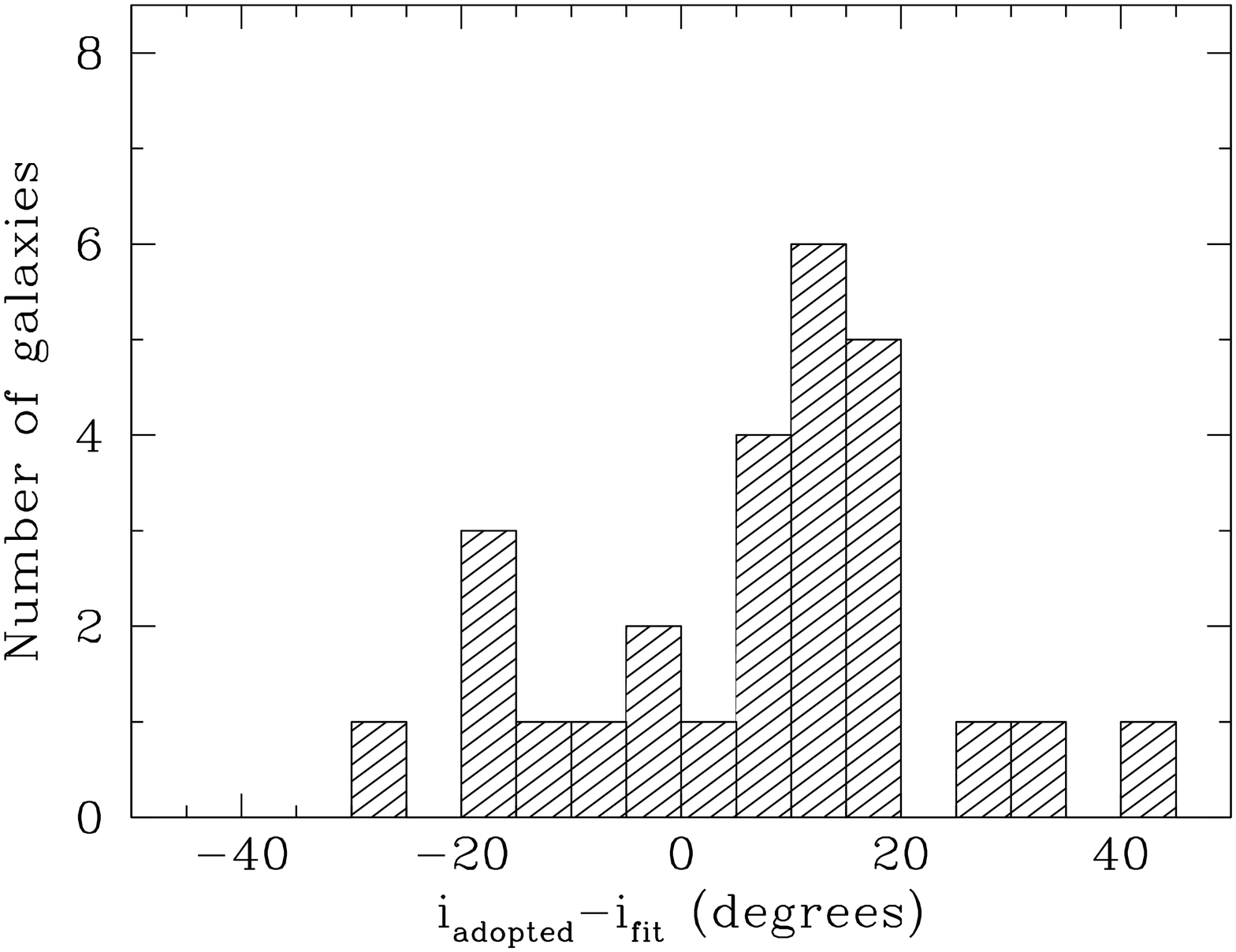}
\caption{Histogram of inclination changes derived from fits to the rotation
  curve with inclination as a free parameter.\label{figinclhist}}
\end{figure}

In Figure~\ref{figinclhist} we plot the histogram of inclination
changes, defined as $i-i_\mathrm{fit}$.  The inclination changes span
a wide range, from $-30^\circ$ (more edge-on in MOND) to $+40^\circ$
(more face-on in MOND). For most galaxies the inclination changes are
positive, which might be explained through bars.  Note that in some
cases the differences between the MOND and the adopted inclinations
may be substantial, which means that for these fits the method used to
fit the inclination may break down because the change in inclination
is large enough to make the rotation curve shape change as well.
However, we are mainly interested in investigating the general trends
and not to get a measurement of the inclination change in this fit.
If the best fit MOND inclination change for a particular galaxy is
large, this likely indicates that MOND cannot be made compatible with
that rotation curve by changing the galaxy's inclination, because for
most galaxies the uncertainty in the inclinations are less than
$10^\circ$.

\begin{figure}
\plotone{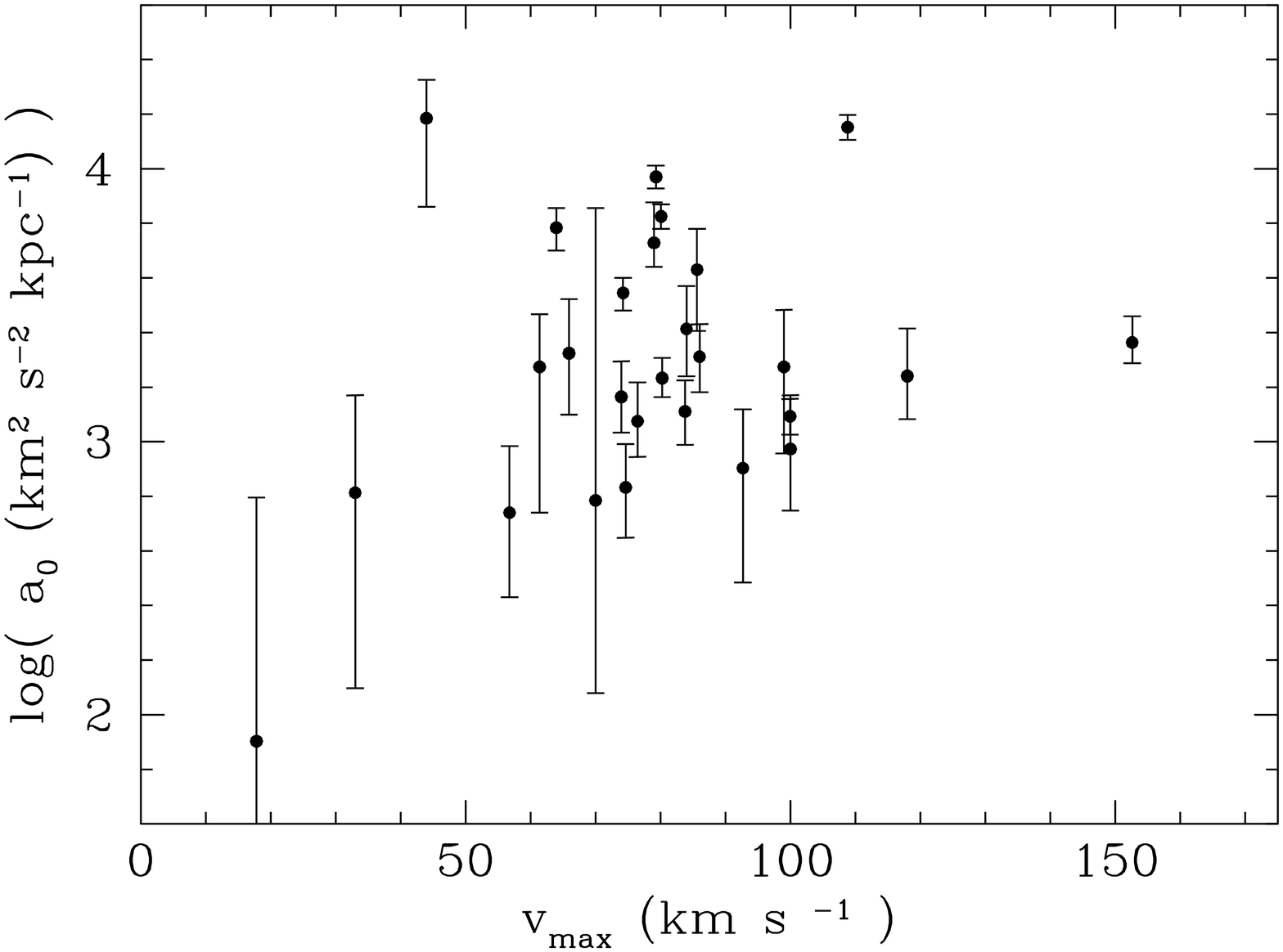}
\caption{Plot of the best fit acceleration parameter $a_0$ versus the maximum
rotation velocity $\varv_\mathrm{max}$. The correlation between these
two parameters as reported by Lake (1989) is not found for the sample
presented here.\label{figa0vsvmax}}
\end{figure}

\subsection{Fits with $a_0$ free}

\noindent Because $a_0$ is a universal parameter in MOND, the MOND
fits should be made with $a_0$ fixed.  However, leaving $a_0$ free
provides both a means to measure $a_0$ (e.g., Begeman et al.\ 1991;
Sanders \& McGaugh 2002), and a way to test MOND. If $a_0$ is truly
universal, it should not depend on galaxy properties. To test this, we
have made fits with $a_0$ as a free parameter. The best fit $a_0$ for
each galaxy is listed in Table~\ref{tabfits}, along with an estimate
of the uncertainties, for which the same caveat as mentioned above
apply.  The best fits are not shown separately in Fig.~\ref{figfits},
because these fits are virtually indistinguishable from the fits with
$d$ as a free parameter.

Lake (1989) reported that the acceleration parameter for his free fits
correlated with the amplitude of the rotation curve. In
Fig.~\ref{figa0vsvmax} we have plotted our best fit $a_0$ versus the
maximum rotation velocity $\varv_\mathrm{max}$. We find no evidence
for a correlation between these parameters.

\begin{figure}
\resizebox{1.0\hsize}{!}{\includegraphics{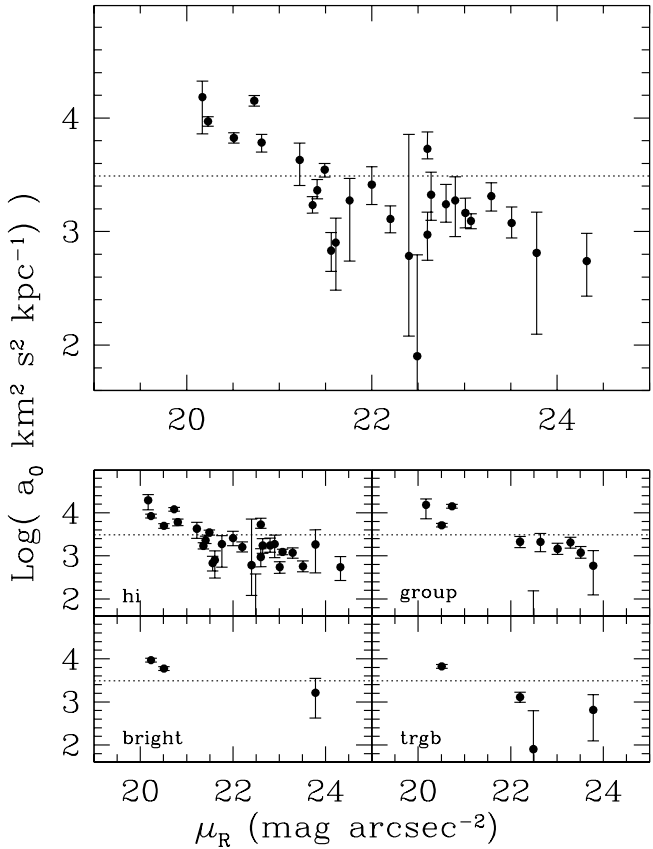}}
\caption{The top panel shows a plot of the best-fit acceleration parameter
$a_0$ against the extrapolated central disk surface brightness
$\mu_R$. There appears to be a correlation between these two
parameters, although this apparent correlation depends strongly on a
small number of points at the high surface brightness end (see
Section~\ref{secdisc}). In the top panel, like in the rest of this
  paper, the adopted distances listed in Table~\ref{tabfits} are used
  in the fits. However, for many galaxies more than one distance
  estimate is available (see e.g., Swaters \& Balcells 2002). In the
  bottom panel, the same correlation as in the top panel is shown for
  each four different methods of distance determination (Hubble flow,
  group membership, brightest star, and tip of the red giant
  branch). It is clear that the apparent trend does not depend on the
  method used to measure the distance.\label{figa0vsmu}}
\end{figure}

Interestingly, as can be seen in Fig.~\ref{figa0vsmu}, there does
appear to be a correlation between $a_0$ and surface brightness, in
the sense that in galaxies with lower surface brightness lower values
for $a_0$ are found.  Taken at face value, the slope of the best fit
to these points is -0.32, i.e., $a_0$ drops by a factor of two for
each magnitude the surface brightness gets fainter. As can be seen in
the bottom panels of Fig.~\ref{figa0vsmu}, this trend does not depend
on the method with which the distance was determined (for galaxies
with different distance estimates we made multiple fits, but elsewhere
only the fit results for the adopted distances as listed in
Table~\ref{tabfits} are presented).

The apparent trend between the acceleration parameter and the central
disk surface brightness depends strongly on the galaxies brighter than
$\mu_R=21$~mag~arcsec$^{-2}$.  We cannot rule out that this trend is
the result of uncertainties (see sections~\ref{secuncert}
and~\ref{secdisc}).

\section{Remarks on individual galaxies}
\label{secindiv}

\noindent Below we discuss the fits shown in
Fig.~\ref{figfits}. Unless otherwise indicated, the `MOND curve'
refers to the fit with only $\Upsilon_{\ast,d}$ as a free parameter.

{\em UGC~731.}--- Swaters \etal\ (1999) showed that this galaxy is
lopsided in its kinematics, with a rotation curve that rises more
steeply on one side of the galaxy than on the other. The average
rotation curve, as presented in Figure~\ref{figfits} is in good
agreement with MOND, except that in the outer parts MOND predicts
somewhat higher rotation velocities.

{\em UGC~3371.}--- The rotation curve of this galaxy has been derived
from a high-resolution \Halpha\ velocity field (see Swaters
\etal\ 2003b). The MOND curve is in good agreement with the observed
rotation curve, although it is slightly above the outer few points.

{\em UGC~4173.}--- The MOND curve based on the \HI\ alone already
predicts significantly higher rotation velocities than are observed.
This galaxy has an optical bar, with a faint surrounding disk, making
the inclination difficult to determine. An inclination of 25$^\circ$
to 30$^\circ$ is consistent with the \HI\ morphology, and would make
this galaxy consistent with MOND, and also bring it closer to the
Tully-Fisher relation.

{\em UGC~4325.}--- The MOND curve and the data agree fairly well. The
MOND curve falls slightly above the observed points in the outer
parts.

{\em UGC~4499.}--- There is good general agreement between the MOND
prediction and the observed rotation curve, although there is
considerable scatter in derived rotation velocities near the center.

{\em UGC~5005.}--- There is good agreement between the data and the
MOND curve. 

{\em UGC~5414.}--- The MOND curve is in excellent agreement with the
rotation curve.

{\em UGC~5721.}--- For this compact, high-surface brightness galaxy,
the MOND curve fails to describe the observed rotation curve. Even
with the inclination or the distance as a free parameter, no good fit
can be found. There is evidence of twisting isophotes in the central
regions (see Swaters \& Balcells 2002), perhaps indicating the
presence of a bar. If so, the associated noncircular motions could
have affected the derived rotation curve.

{\em UGC~5750.}--- The MOND curve is in good general agreement with the 
observed rotation curve.

{\em UGC~6446}--- The MOND curve falls below the observed rotation
curve in the central parts, and above the rotation curve in the
outer parts.  If this galaxy is part of the Ursa Major cluster (e.g.,
Tully et al.~1996), the increased distance of 18.6~Mpc (Tully \&
Pierce 2000) would make this galaxy less compatible with MOND.
However, this galaxy is on the edge of the boundary taken to define
the UMa cluster in both velocity and angle on the sky, so it may not
be a genuine cluster member. In this paper, we have adopted a distance
of 12.8~Mpc.

{\em UGC~7232.}--- The MOND curve and the data agree fairly well.

{\em UGC~7323.}--- The MOND curve is somewhat higher than the observed
curve in the central parts, but there is good general agreement.

{\em UGC~7399.}--- For this compact galaxy, the MOND curve does not
agree with the observed rotation curve, as it predicts a much steeper
central gradient than is observed in the rotation curve. The bar in
this galaxy may have affected the derived rotation curve.

{\em UGC~7524.}--- The MOND curve and the observed rotation curve for
this well-resolved galaxy agree well, except in the outer points. 

{\em UGC~7559.}--- The MOND curve agrees well with the observed curve
in the inner parts, but it predicts significantly higher-than-observed
rotation velocities in the outer parts. Like UGC~5721, this galaxy
appears to have a bar, and the associated noncircular motions could
have affected the derived rotation curve.

{\em UGC~7577.}--- The observed rotation curve of this galaxy can be
explained by gas and stars alone, even with Newtonian
gravity. Consequently, the MOND curve predicts significantly higher
rotation velocities in the outer parts. A lower distance and a lower
inclination would help to make the MOND prediction more compatible
with the observed curve.

{\em UGC~7603.}--- Although the MOND curve falls somewhat below the
observed rotation curve in the outer parts, there is good general
agreement.

{\em UGC~8490.}--- The MOND prediction for this galaxy falls
systematically below the observed flat part of this rotation curve. A
larger distance makes this rotation curve more consistent with the
MOND curve, although the MOND curve still falls short of the observed
rotation curve near the turnover in the rotation curve. Because of the
large warp in this galaxy, the shape and amplitude of the rotation
curve are uncertain.

{\em UGC~9211.}--- The curve predicted by MOND and the observed curve
are in good agreement. 

{\em UGC~11707.}--- The rotation curve predicted by MOND falls below
the inner rotation curve, and above the outer rotation curve.

{\em UGC~11861.}--- The MOND curve and the observations are in good
agreement.

{\em UGC~12060.}--- The rotation curve as predicted by MOND falls
below the observed one in the inner parts of the rotation curve, and
is above the observed one in the outer parts. An inclination of
$20^\circ$ in stead of $40^\circ$ is required for agreement between
MOND and the observations. Such a low inclination is consistent with
the optical morphology of this galaxy, but appears inconsistent with
the Tully-Fisher relation.

{\em UGC~12632.}--- The rotation curve predicted by MOND falls below
the observed curve in the inner parts and above the observed curve in
the outer parts.

{\em F568-V1.}--- There is good agreement between the MOND prediction
and the observed rotation curve.

{\em F574-1.}--- Except for the one outermost point in the MOND
prediction, which may well be due to uncertainties in the contribution
of the \HI\ to rotation curve, the MOND curve and the observed one are
in good agreement.

{\em F583-1.}--- Although the MOND curve falls slightly above the
observed velocities in the center, and slightly below at intermediate
radii, there is good general agreement between the observed rotation
curve and the MOND prediction.

{\em F583-4.}--- There is good agreement between the observed and the
MOND predicted rotation curves.

\section{Uncertainties}
\label{secuncert}

\noindent As is clear from the notes on individual galaxies and the
fits shown in Fig.~\ref{figfits}, for about a quarter of the galaxies
in the sample there are noticeable discrepancies between the observed
and the predicted curves, and in a few instances there are significant
inconsistencies. On the one hand, these inconsistencies may signal a
problem for MOND, but, on the other, they might be the result of
uncertainties associated with the galaxies in this sample.  Because
MOND fits are essentially one-parameter fits, with only the
mass-to-light ratio of the stars as a free parameter, it is much more
sensitive to observational uncertainties than fits with dark matter
halos. For such fits with dark matter halos, any uncertainties in the
rotation curve or in the contribution of the stars and gas, for
example as a result of uncertainties in distance or inclination, can
be accommodated by changing the parameters of the dark matter
halo. The MOND fits, which are directly tied to the distribution of
the gas and stars, cannot accommodate these uncertainties. As a
result, uncertainties in e.g., the rotation curves, the distances, or
the inclinations can lead to apparently good dark matter fits
(although likely with incorrect halo parameters), whereas the same
uncertainties can lead to very poor MOND predictions of the rotation
curves.  It is therefore imperative to consider the uncertainties that
may play a role for the data presented here before we can discuss the
implications of our results for MOND.

\subsection{Observational uncertainties}

\noindent Given that the sample presented here consist in large part
of nearby dwarf galaxies, uncertainties in distances will likely play
a role. For those nearby galaxies, the distances have been determined
with different methods: Hubble flow, group membership, brightest
stars, and tip of the red giant branch. Some of these methods can
produce uncertain distance measurements. As we have shown in this
paper, the MOND fits can be improved by adopting different
distances. It is likely that in individual cases incorrect distance
estimates have contributed to poor MOND predictions of the rotation
curves. However, it is unlikely that changes in distance alone can
explain all the poor predictions, because that would require the
galaxies in our sample to be preferentially closer, even galaxies at
large distances which presumably have more accurate distances.

Another important factor that may contribute to uncertainties in the
MOND fits are the inclinations of the galaxies in our sample. As
described in Section~\ref{incfits}, many of the galaxies in our sample
have irregular appearances and slowly rising rotation curves, making
it difficult to determine their inclinations from either their
morphology or their kinematics. Moreover, towards more face-on
galaxies, intrinsic noncircular shapes may lead to an overestimate of
the inclination, and towards higher inclinations the uncertainties in
the intrinsic thicknesses of these galaxies may affect the derived
inclinations. Thus, the inclinations of the galaxies in our sample may
be uncertain.

We have shown that the MOND fits can significantly be improved by
adopting different inclinations. Moreover, for some galaxies, like
UGC~4173 and UGC~12060, the MOND-preferred inclinations not only
improve the MOND fits, but also are compatible with the galaxies
morphology and kinematics. In the case of UGC~4173, the MOND-preferred
inclination also brings the galaxy closer to the Tully-Fisher
relation. However, for most galaxies the required changes in
inclination are larger than the expected uncertainties. In addition,
if changes in inclination would be the main cause of the poor MOND
predictions of the rotation curves, then the majority of the galaxies
in our sample would have to be more face-on than the inclinations
reported in Table~\ref{tabsample}. This could be possible if the
galaxies in our sample have significant intrinsic ellipticities or
strong bars. However, to explain differences of up to $20^\circ$ even
at inclinations of about $50^\circ$, we estimate an intrinsic
ellipticity of 0.35 is needed. Such large intrinsic ellipticities are
not seen among late-type disk galaxies (e.g., Schoenmakers et
al.\ 1997; Andersen et al.\ 2001), although they might occur in low
luminosity systems (Sung et al.\ 1998). However, there is little
evidence of the perturbations in the velocity fields associated with
such strong ellipticities (Schoenmakers et al.\ 1997) in the velocity
fields presented in Swaters et al.\ (2002).  It is therefore unlikely
that all the poor MOND predictions can be explained by uncertainties
in the inclination alone, but it is likely that in some individual
cases incorrect adopted inclinations can be the main reason for poor
MOND predictions.

Naturally, it is also possible that both the adopted inclination and
distance are incorrect. We investigated this in our fitting process
but found that in all but a few of our fits, distance and inclination
are generally strongly covariant, and in most cases it was not
possible to constrain both of these parameters simultaneously. Still,
when compared to leaving only the distance or the inclination free in
the fits, if both are left free, on average smaller changes are
required to make the MOND-predicted curve more compatible with
observed rotation curve. At the same time, the required changes are
still systematic, i.e., on average the galaxies still need to be
closer and more face-on.

In addition to the uncertainties in distance and inclination, other
factors may also play a role. These include beam smearing, noncircular
motions, morphological asymmetries, corrections for asymmetric drift,
and uncertainties in the photometric calibration.  Even though the
\HI\ rotation curves presented here have been corrected for beam
smearing (see SSAH), the inner rotation curve shape may be uncertain
simply due to lack of resolution. For many galaxies in our sample
high-resolution \Halpha\ rotation curves are available to mitigate
this effect, but \Halpha\ rotation curves themselves may be uncertain
because they are determined along a one-dimensional slice and not from
the velocity field as a whole and hence are more susceptible to
noncircular motions. In addition, the rotation curves presented here
have not been corrected for asymmetric drift because it was deemed not
to be important. Another potential factor are morphological and
kinematical asymmetries. Not only can these affect the derived baryon
distributions and rotation curves, they are also incompatible with the
underlying assumption of axisymmetry used in calculating the
MOND-predicted rotation curves. All these factors combined could
result in a derived rotation curve that is not representative of the
true circular velocity.

\subsection{Fitting uncertainties}

\noindent In addition to observational uncertainties, the quality of
the fits may be affected by the fitting process and the associated
assumptions.  An example is the adopted thickness for the \HI\ and
stellar disks.  Here, for ease of comparison with other studies, we
have assumed that the gaseous and stellar disks are infinitely
thin. Compared to a stellar disk that is assumed to have a scale
height of 0.2 disk scale lengths, as was adopted by Swaters (1999),
the rotation curve calculated for an infinitely thin disk rises more
steeply, on average by 15\%, but reaching up to 40\% for galaxies with
central concentrations of light. The amplitude in the thin disk case
is about 5-10\% higher than in the thick disk case. Thus, taking a
thick disk in stead of a thin disk would results in M/Ls that are 10\%
to 20\% higher, and values for $a_0$ that are correspondingly lower.
The rotation curve of the \HI\ is less sensitive to the adopted
thickness. The difference between an infinitely thin disk and a disk
with a scale height of 0.2 optical disk scale lengths is in general
well below 5\%, and hence the choice of thickness of the \HI\ disk has
little influence on the fits.

Another uncertainty on the results is the adopted form of $\mu$ (see
Eq.~\ref{eqmond}), that describes the transition between the Newtonian
and MOND regimes. In the results reported in the Tables and Figures in
this paper, we have used the commonly used form given in
Eq.~\ref{eqmu}. We have also tried the form:
\begin{equation}
\mu(x) = x/(1+x),
\label{eqmu2}
\end{equation}
as used by Famaey \& Binney (2005) and McGaugh (2008). This revised
transition formula results in more noticeable deviations from
Newtonian gravity at higher accelerations.  This has an effect on the
fits, and we looked in particular in terms of the possible correlation
between $a_0$ and surface brightness as shown in
Fig.~\ref{figa0vsmu}. When Eq.~\ref{eqmu2} is used, galaxies that are
deeper in the MOND regime (i.e., the LSB galaxies) tend to end up with
lower values for $a_0$, whereas the derived values for $a_0$ change
little for high surface brightness galaxies.

The fact that some of the rotation curves in our sample have been
derived from a combination of \HI\ and \Halpha\ data might have
affected our fit results, because the long-slit \Halpha\ data sample
only a single slice through the velocity field, making them more
sensitive to the effects of noncircular motions. In addition, the
higher spatial sampling in the galaxy centers might bias the
fits. However, any such bias would, at least partly, be offset by the
usually larger uncertainties on the \Halpha\ rotation velocities. Even
though such a bias could have played a role, we find it does not for
the galaxies in our sample. For all galaxies for which we have hybrid
\HI-\Halpha\ rotation curves, we have made fits to the \HI-only
rotation curves and found no significant differences. This is also
illustrated by the fact that there are no systematic differences
between the galaxies in our sample for which we have \HI-only rotation
curves, and those for which we have hybrid rotation curves.

Perhaps the biggest uncertainties on the derived parameters are from
the fitting process itself. The values reported in Table~\ref{tabfits}
are best-fit parameters. However, in some cases, the best-fit values
suggest large changes in distance, inclination, or $a_0$, even though
the nominal values provide fits that are nearly as good in a $\chi^2$
sense (e.g, UGC~5414 and F583-4). In most of these cases, the
estimated uncertainties on the derived parameters are large, thereby
indicating that a large range in parameter space produces fits of
similar quality. In a number of cases, however, the best-fit values
for the derived parameters are off from the nominal value by a large
amount, even when a visual comparison of the MOND fits and the fits
with additional free parameters suggests that both fits are comparable
in quality. Examples of this are UGC~3371, UGC~11861, and F574-1. In
light of the uncertainties on the rotation velocity uncertainties, as
was also mentioned in Sec.~\ref{secfitres}, the best-fit values
derived from the fits should not be taken at face value.

\section{Discussion}
\label{secdisc}

\noindent We have found that the rotation curves as predicted by MOND
for the sample of dwarf and LSB galaxies presented in this paper are
generally in good agreement with the observed rotation curves for
roughly three quarters of the sample. This is remarkable, given that
MOND is a one parameter fit with only M/L as a free parameter. It is
even more remarkable, in light of the uncertainties associated with a
sample of low surface brightness and dwarf galaxies, as was discussed
in more detail in the previous section. Given these uncertainties, a
fraction of around a quarter of galaxies for which MOND does not
adequately predict the observed rotation curves does not seem
unexpected, and hence the differences between the observed rotation
curves and the MOND fits may not signal a failure of MOND, but rather
reflect the uncertainties associated with the galaxies in this sample.

Despite these uncertainties, there are a number of interesting results
from our study. One is that the MOND fits appear to deviate in a
systematic way from the observed rotation curves. In almost 80\% of
the galaxies (21 out of 27) the MOND curve predicts higher rotation
velocities in the outer parts than are observed. In 40\% of the
galaxies in our sample, MOND predicts rotation velocities that are
higher by more than twice the uncertainty in the rotation velocity of
the last measured point of the rotation curve.  As we have shown,
these discrepancies can be explained if these galaxies have different
inclinations or distances. To explain the apparent systematic
deviations between the MOND curves and the observed rotation curves in
terms of changes in distance or inclination, the galaxies in our
sample need to be preferentially closer, need to have preferentially
lower inclinations, or both, which seems unlikely.

An alternative explanation for the fact that most MOND-predicted
curves tend to have higher rotation velocities in the outer parts than
is observed, is that the value for $a_0$ is preferentially lower. Of
course, in MOND $a_0$ is a universal constant and should not vary from
galaxy to galaxy. In the MOND fits presented here, we have adopted
$a_0=3080$ km$^2$ s$^{-2}$ kpc$^{-1}$ (e.g., Sanders \& McGaugh
2002). If we average the $a_0$ derived from our fits with the
acceleration parameter free, we find $a_0=3350$ km$^2$ s$^{-2}$
kpc$^{-1}$, consistent with the value used by Sanders \& McGaugh
(2002). However, as can be seen in \ref{figa0vsmu}, most of our fits
result in low values for $a_0$. Excluding the three highest values for
$a_0$, we find $a_0=2150$ km$^2$ s$^{-2}$ kpc$^{-1}$. Thus, a somewhat
lower value for $a_0$ would, on average, lead to improved fits for the
galaxies in our sample.

We also have found what appears to be a correlation between $a_0$ and
surface brightness. Such a trend, if real, would be a problem for
MOND, because $a_0$ is a universal constant. However, it seems that
this correlation is largely caused by a small number of galaxies at
the high surface brightness end, for which large values for $a_0$ were
found: UGC~5721, UGC 7232, UGC~7399, UGC~7603, and UGC~8490.  As we
noted in Sec.~\ref{secindiv}, UGC~5721 and UGC~7399 are compact
galaxies with a central bar, and hence the inner rotation curve could
be affected by noncircular motions. UGC~8490 has a strong warp,
making the outer rotation curve less certain. UGC~7232 only has four
points in its rotation curve and large uncertainties on the fitted
value of $a_0$. Considering these factors, it is possible that the apparent
correlation is the result of uncertainties, and not of a true
underlying relation.

Previous studies of MOND rotation curves that included LSB galaxies in
their samples have not reported a trend between surface brightness and
$a_0$.  De Blok \& McGaugh (1998) analyzed the \HI\ rotation curves
presented in de Blok et al.\ (1996) in the context of MOND, and found
most galaxies to be consistent with MOND, although for a few
adjustments to distance and inclination had to be made. The sample
presented in Sanders \& Verheijen (1998) contained 12 LSB galaxies,
but the authors did not report a trend between $a_0$ and surface
brightness. Neither does an inspection of the LSB galaxies in their
sample show the prediction by MOND of systematically faster rotation in
the outer rotation curves, like is seen in our sample, except for
UGC~6446, which is a galaxy also included in this sample. However, our
sample extends more than a magnitude deeper in surface brightness,
making the sample presented here more sensitive to effects that may
play a role at low surface brightness.

Interestingly, it may be possible to explain, within the context of
MOND, why at least some galaxies would appear to have a low value for
$a_0$ when $a_0$ is left as a free parameter in rotation curve
fits. In the results presented here, we have only considered the
accelerations within the galaxies themselves. However, whether an
object is in the MOND regime is determined by the overall
acceleration, which includes external effects (Milgrom 1983a,b). Thus,
if an object, with internal accelerations that would place it in the
MOND regime, is placed in a Newtonian external acceleration that is
larger than $a_0$, its kinematics would be entirely Newtonian. If one
were to fit $a_0$ in that case, one would derive a value of zero,
because the object would never enter the MOND regime. Of course, the
effects of the external fields are much smaller for the galaxies in
our sample. Even for the galaxy in our sample with the closest
neighbor (UGC~7577, near NGC~4449), the external acceleration is
estimated to be $1\times 10^{-9}$ cm s$^{-2}$ (assuming a projected
distance of 14.4 Mpc). Even though the amplitude of the external field
is difficult to determine because it depends on the details of the
galaxy environment, it can only lower the value of $a_0$ in fits where
$a_0$ is left free. Moreover, any external fields will have the
strongest effects in LSB galaxies, where the internal accelerations
are lowest. Thus, while external fields may contribute to the apparent
correlation between surface brightness and $a_0$, it is difficult to
establish this quantitatively.

Given the uncertainties in the distances, the inclinations, other
observational uncertainties, uncertainties in the
fitting process, and possible effects of external
  fields, it seems that the correlation between surface brightness
and $a_0$ may not be significant. Perhaps more certain is the finding
that the galaxies in the sample presented here suggest a lower value
for $a_0$ than the adopted value of $3080$ km$^2$ s$^{-2}$ kpc$^{-1}$.

To thoroughly test MOND in the LSB and dwarf galaxy regime, galaxies
are needed with reliable distance measurements and inclinations,
well-resolved rotation curves and symmetric appearance, both in
morphology and kinematics. None of the galaxies in our sample meet all
these criteria simultaneously.  However, there are four galaxies that
meet all but one: UGC~3371, UGC~6446, UGC~7524, and
UGC~12632. UGC~7524 is the only one with a reliable distance
measurement from the TRGB (Karachentsev et al.\ 2003), but it has
asymmetric kinematics (Swaters et al.\ 1999).

\section{Summary and conclusions}
\label{secconc}

\noindent We have presented MOND fits for a sample of 27 dwarf and LSB
galaxies.  MOND is remarkably successful at predicting the general
shape of the rotation curves in this sample. Only for approximately a
quarter of the galaxies MOND does not adequately predict the observed
rotation curves.  This is remarkable given the uncertainties
associated with the galaxies in this sample, especially the
uncertainties in distance and inclination. Close inspection of the
fits indicates that for almost 80\% of the galaxies MOND predicts
higher velocities in the outer rotation curves than is observed.
Although this could be the result of uncertainties, this systematic
deviation can also be explained if the MOND acceleration parameter is
slightly lower than usually assumed, with a value of $a_0=2150$ km$^2$
s$^{-2}$ kpc$^{-1}$ (or $0.7 \times 10^{-8}$ cm~s$^{-2}$). We find
that there appears to be some evidence of a correlation between $a_0$
and the central extrapolated disk surface brightness $\mu_R$, but this
possible correlation depends heavily on a few galaxies at high surface
brightness, whose rotation curves may be uncertain because of bars or
warps.  At the low surface brightness end, the derived values of $a_0$
could be be biased towards low values if external fields are
important.  Overall, the uncertainties associated with this sample
make it difficult to draw strong conclusions. Improved distances and
inclination estimates for these galaxies could make it possible to
test MOND more strongly.

\end{document}